\documentstyle[psfig]{mn2e}

\begin{document}

\title[Origin of Metallicity Dependence of Exoplanets]
  {Origin of the Metallicity Dependence of Exoplanet Host Stars
  in the Protoplanetary Disk Mass Distribution}

\author[M. C. Wyatt et al.]
  {M. C. Wyatt$^1$\thanks{Email: wyatt@ast.cam.ac.uk}, C. J. Clarke$^1$
  and J. S. Greaves$^2$ \\
  $^1$ Institute of Astronomy, University of Cambridge, Madingley Road,
  Cambridge CB3 0HA, UK \\
  $^2$ Scottish Universities Physics Alliance, University of St. Andrews,
  Physics \& Astronomy, North Haugh, St Andrews KY16 9SS, UK}

\maketitle

\begin{abstract}
The probability of a star hosting a planet that is detectable
in radial velocity surveys increases $P_{\rm{pl}}(Z) \propto (10^Z)^2$,
where $Z$ is stellar metallicity.
Models of planet formation by core accretion
reproduce this trend, since the protoplanetary disk of a high
metallicity star has a high density of solids and so forms planetary cores
which accrete gas before the primordial gas disk dissipates.
This paper considers the origin of the form of the metallicity dependence
of $P_{\rm{pl}}(Z)$.
We introduce a simple model in which detectable planets form when the
mass of solid material in the protoplanetary disk, $M_{\rm{s}}$, exceeds a 
critical value.
In this model the form of $P_{\rm{pl}}(Z)$ is a direct reflection of the
distribution of protoplanetary disk masses, $M_{\rm{g}}$, and the observed
metallicity relation is reproduced if
$P(M_{\rm{g}}>M_{\rm{g}}^{'}) \propto (M_{\rm{g}}^{'})^{-2}$.
We argue that a protoplanetary disk's dust mass measured in sub-mm observations
is a relatively pristine indicator of the mass available for planet-building and
find that the disk mass distribution derived from such observations is consistent
with the observed $P_{\rm{pl}}(Z)$ if a solid mass $M_{\rm{s}}>0.5M_{\rm{J}}$
is required to form detectable planets.
Any planet formation model which imposes a critical solid mass for detectable planets
to form would reproduce the observed metallicity relation, and core
accretion models are empirically consistent with such a threshold criterion.
While the outcome of planet formation in individual systems is
debatable, we identify 7 protoplanetary disks which, by rigid application of
this criterion, would be expected to form detectable planets and may provide
insight into the physical conditions required to form such planets.
A testable prediction of the model is that the metallicity dependence should
flatten both for $Z>0.5$ dex and as more distant and lower mass planets are
discovered.
Further, combining this model with one in which the evolution of a star's debris 
disk is also influenced by the solid mass in its protoplanetary disk, results in the 
prediction that debris disks detected around stars with planets should be
more infrared luminous than those around stars without planets in tentative
agreement with recent observations.
\end{abstract}


\section{Introduction}
\label{s:intro}
The study of how planetary systems form and evolve
was revolutionised when the first extrasolar
planet was discovered in radial velocity studies of
the star 51 Peg (Mayor \& Queloz 1995).
Over 200 extrasolar planets are now known
(Butler et al. 2006), and studying these planets has
yielded enormous advances in our understanding of how
they formed (Papaloizou \& Terquem 2006; Udry et al. 2007).
Perhaps the most telling discovery was that of a correlation
in the probability of a star hosting a planet, $P_{\rm{pl}}$,
which is found to increase with stellar metallicity
(Gonzalez 1997).
Fischer \& Valenti (2005; hereafter FV05) found that, for
stars with a metallicity $Z=$ [Fe/H] between -0.5 and 0.5 dex,
the metallicity dependence of the fraction of stars with
planets with orbital periods $<4$ years and with amplitudes
in radial velocity studies in excess of $K>30$ m s$^{-1}$
(i.e., Saturn-Jupiter mass planets, depending on orbital
period) is
\begin{equation}
  P_{\rm{pl}}(Z) = 0.03 \times 10^{2Z},
  \label{eq:pplfv05}
\end{equation}
which corresponds to a planet fraction which increases with
the square of the number of iron atoms in the stellar atmosphere.
Similar trends have been found to apply to all species including
Si and Ni (e.g., Ecuvillon et al. 2004; Robinson et al. 2006;
Gonzalez 2006).
The origin of this metallicity dependence is thought to be
intrinsic to the planet formation process (FV05), and not
caused by contamination from planetesimals falling
onto the star, as is believed to be the cause of the high metallicities
of DAZ white dwarfs (Jura 2006; Kilic \& Redfield 2007),
although the recent discovery that planet hosting giant stars
do not favour metal rich systems is currently reigniting this debate
(Pasquini et al. 2007).

Since the discovery of the extrasolar planet metallicity correlation,
much work has gone into considering how stellar metallicity could affect
different aspects of the planet formation process in the various
models (e.g., Livio \& Pringle 2003).
It has been found that forming planets by gravitational instability does
not introduce any significant metallicity dependence (Boss 2002; Cai et al. 2006),
whereas models of planet formation by core accretion seem to
readily reproduce the observed trend (Ida \& Lin 2004b; Kornet et al. 2005;
Benz et al. 2006; Robinson et al. 2006).
This is because, in the core accretion models, planetesimals grow into planet cores
through collisions, subsequently accreting gas from the surrounding gas
disk once they become large enough, and then interacting with that disk so
as to migrate inward (e.g., Lin \& Papaloizou 1986; Papaloizou et al. 2007).
The core accretion models predict a metallicity dependence because
a higher metallicity implies higher solid mass and hence faster core growth,
which means that the critical core mass for gas accretion can occur
before the gas disk dissipates on $\sim 6$ Myr timescales (Haisch, Lada
\& Lada 2001; Clarke, Gendrin \& Sotomayer 2001).
However, it remains to be explained why the metallicity dependence
has a form $\propto 10^{2Z}$ as opposed to, e.g.,
$\propto 10^Z$.
The origin of the dependence found in these models is hidden somewhere
within the large number of model components of which they are comprised,
although it has been shown that a large solid disk mass is required
if planets are to form (Ida \& Lin 2004b).

In this paper we consider the origin of the form of the metallicity
dependence using a simple heuristic model in which detectable planets
form as long as the solid mass of material in the protoplanetary disk
exceeds a critical value (e.g., Greaves et al. 2007).
That model is described in \S \ref{s:model}, where it is shown how
the metallicity relation is then directly related to the initial
disk mass distribution.
This section also compares the disk mass distribution required to
reproduce the observed planet-metallicity trend in this model
with that inferred from sub-mm observations of star forming regions.
The implications of this model are discussed in \S \ref{s:disc}, along
with a discussion of why the solid mass should provide such a
strong constraint on whether a system goes on to form a detectable planet.
The conclusions are given in \S \ref{s:conc}.

\section{Critical solid mass model}
\label{s:model}
This model assumes that stars form surrounded by a protoplanetary
disk which is made up of both solids and gas.
We denote the mass of each of these components by $M_{\rm{s}}$
and $M_{\rm{g}}$, respectively.
The gaseous component dominates the total mass of the disk, and it
is assumed that the outcome of the star formation process results
in some universal distribution of disk masses (i.e., gas masses),
which we define by the probability of any given star having
had a protoplanetary disk with a gas mass larger than 
$M_{\rm{g}}^{'}$ as $P(M_{\rm{g}}>M_{\rm{g}}^{'})$.
The solid mass of any given disk is assumed to be directly
related to the mass of the gaseous component through the
final metallicity of the star (e.g., Greaves et al. 2007):
\begin{equation}
  M_{\rm{s}} = 0.01 M_{\rm{g}} 10^Z.
  \label{eq:ms}
\end{equation}
Here we have assumed that the ratio of gas to
solids is 100 for stars formed in a $Z=0$
environment, consistent with that seen in nearby star
forming regions (James et al. 2006).
Thus it is assumed that stellar metallicities are indicative of
the conditions present prior to the formation of the star that
continued to be reflected in the composition of the
protoplanetary disk, and that exerted no influence over the
resulting distribution of protoplanetary disk
masses.

The most important assumption is then that all of the stars
that have disks with $M_{\rm{s}}$ larger than some critical
value $M_{\rm{s,crit}}$ go on to form planets which can be
detected in radial velocity surveys, i.e.,
$P_{\rm{pl}} = P(M_{\rm{s}}>M_{\rm{s,crit}})$.
The physical origin for this critical value is not part of
this heuristic model, although it does have a physical motivation
based on core accretion models (e.g., Ida \& Lin 2004b), as
discussed in \S \ref{s:intro} and in more detail in \S \ref{s:disc}.

\subsection{Analytical solution}
\label{ss:analyt}
Since the probability of forming a planet depends only on the
solid mass, the critical mass above which the total disk mass
(i.e., gas mass) must be to form a planet is dependent on
metallicity:
\begin{equation}
  M_{\rm{g, crit}} = 100 M_{\rm{s, crit}} 10^{-Z}.
  \label{eq:mgcrit}
\end{equation} 
The gas mass distribution is assumed to be independent of
metallicity, and so the probability of any star forming
a planet is metallicity dependent, since $P_{\rm{pl}} = 
P(M_{\rm{g}}>M_{\rm{g,crit}})$.
Thus to reproduce equation (\ref{eq:pplfv05}) requires
a gas mass distribution in which:
\begin{equation}
  P(M_{\rm{g}} > M_{\rm{g}}^{'}) =
    0.03 (100 M_{\rm{s,crit}}/M_{\rm{g}}^{'})^2,
  \label{eq:pmgan}
\end{equation}
where the critical solid mass required to form a planet, $M_{\rm{s,crit}}$,
is some as yet undefined constant.
Since the probability of any star hosting a planet given
in equation (\ref{eq:pplfv05}) is only known to apply for
$P_{\rm{pl}} < 0.25$ (due to the lack of surveys at higher $Z$),
it follows that the distribution given in equation (\ref{eq:pmgan}) 
is also only valid for $P(M_{\rm{g}} > M_{\rm{g}}^{'}) < 0.25$, and
so for $M_{\rm{g}} > \sqrt{1200}M_{\rm{s,crit}}$.
Thus, in this model the observed $P_{\rm{pl}}(Z)$
in equation (\ref{eq:pplfv05}) is telling us about the mass
distribution of the most massive 25\% of disks.

\subsection{Gas disk distribution from observations}
\label{ss:obs}
The gas mass distribution required by this model in order to match the
observed $P_{\rm{pl}}(Z)$ (equation \ref{eq:pmgan})
can now be compared with the observed gas mass distribution.
The gas mass distribution of protoplanetary disks is not well-known
because the majority of that mass is in molecular hydrogen
which is difficult to detect, especially in the cold outer regions of
the disks where most of the mass resides (Thi et al. 2001; Sheret, Ramsay-Howat
\& Dent 2003).
Species such as CO are easier to detect (e.g., Dent, Greaves \&
Coulson 2005; Dutrey, Guilloteau \& Ho 2007), however there is uncertainty in 
the CO/H$_2$ ratio because some of this gas ends up frozen onto dust grains
or photo-dissociated (Dullemond et al. 2007; Najita et al. 2007).
On the other hand, the dust mass distribution of protoplanetary disks is well
characterised, since this can be measured with relatively few uncertainties
from sub-mm and mm wavelength observations (Andr\'{e} \& Montmerle 1994;
Beckwith, Henning, \& Nakagawa 2000).

\begin{figure}
  \begin{center}
    \vspace{-0.2in}
    \begin{tabular}{c}
      \hspace{-0.35in} \psfig{figure=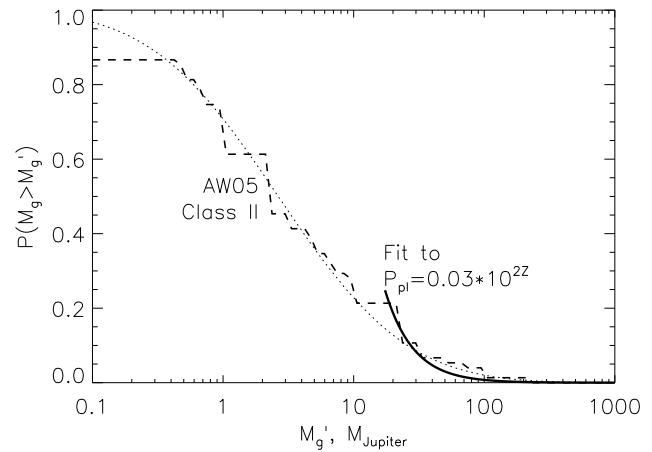,height=2.6in}
    \end{tabular}
    \caption{Distribution of protoplanetary disk gas masses.
    The gas mass distribution inferred from the
    dust mass distribution of class II objects in Taurus-Auriga
    (Andrews \& Williams 2005) is shown with a dashed line.
    The dotted line is a log-normal fit to this distribution
    centred on $2.5M_{\rm{J}}$ of width 0.77 dex.
    The distribution required in the critical solid mass model to fit
    the extrasolar planet metallicity relation (equations 
    \ref{eq:pplfv05} and \ref{eq:pmgan}) is shown with a solid line,
    assuming $M_{\rm{s, crit}}=0.5M_{\rm{J}}$.}
    \label{fig:gasmass}
  \end{center}
\end{figure}

Here we make the assumption that dust mass can be used as a
proxy for the total gas mass in protoplanetary disks (for a fixed $Z$),
and we derive the gas mass distribution from the dust mass distribution in 
Taurus-Auriga,
which was measured using sub-mm photometry of 153 pre-main sequence stars
by Andrews \& Williams (2005; hereafter AW05).
Since the stars in the AW05 sample are at a range
of evolutionary stages we chose to use only the disk masses of the
75 class II objects (i.e., T Tauri stars) in their sample to ensure that
the disk mass distribution is indicative of that at the epoch of planet
formation.
Class I sources were omitted because of a potential
contribution to the sub-mm flux from a remnant circumstellar
envelope.
Class III sources were omitted because of the possibility that their
currently low disk masses are a consequence of the disks being at an
advanced evolutionary stage, and so are not necessarily indicative of
a low mass present at the planet forming epoch.
To obtain the gas mass distribution, the gas/dust ratio was assumed to be 100
for all stars, based on the metallicities in nearby star forming regions being
close to solar with a small dispersion for each region (Padgett 1996; Vuong et
al. 2003; James et al. 2006).
The mass distribution of class II objects is shown in Fig.~\ref{fig:gasmass}.
Ten objects from this sample have only upper limits to their disk masses,
which were set to zero in Fig.~\ref{fig:gasmass}.
Since these upper limits are $\leq 1M_{\rm{J}}$, we infer that
the disk mass distribution is accurate for the most massive 69\%
(52/75) of disks that are above this limit ($\geq 1M_{\rm{J}}$).
\footnote{We note that, even though 10 of the AW05 class II sources
were not detected in individual sub-mm photometry observations, co-addition
of this data-set leads to a net positive detection of $2.7 \pm 0.9$ mJy, 
corresponding to a mean disk mass of $0.14M_{\rm{J}}$ which is consistent
with that expected from the log-normal distribution plotted in 
Fig.~\ref{fig:gasmass}.}

\begin{figure}
  \begin{center}
      \vspace{-0.2in}
    \begin{tabular}{c}
      \hspace{-0.3in} \psfig{figure=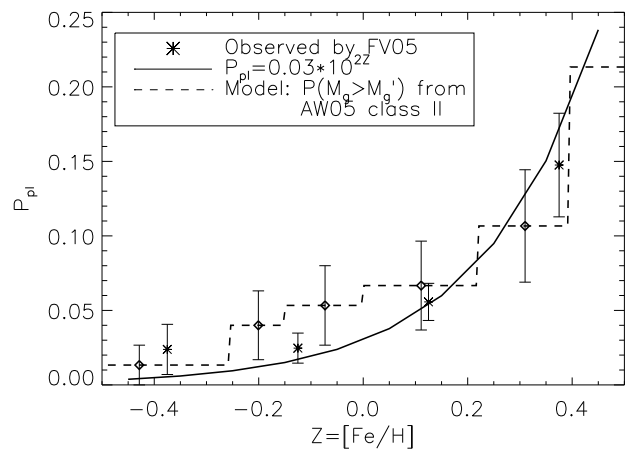,height=2.6in}
    \end{tabular}
    \caption{
    Probability that a star of given metallicity has an extrasolar
    planet that is detectable in the current radial velocity surveys.
    The predictions of the critical solid mass model
    based on the Andrews \& Williams (2005) distribution of dust masses
    of class II protoplanetary disks in Taurus-Auriga is
    shown with a dashed line, with errors indicated by diamonds with
    $\sqrt{N}$ error bars.
    The asterisks shows the results of the radial velocity survey of
    Fischer \& Valenti (2005) with $\sqrt{N}$ error bars.
    The fit to the FV05 data (equation \ref{eq:pplfv05}) is shown with a
    solid line.}
    \label{fig:pplz2}
  \end{center}
\end{figure}

The critical solid mass model (\S \ref{ss:analyt}) was used to
determine the metallicity relation predicted from the observed
gas mass distribution:
\begin{equation}
  P_{\rm{pl}} = [N(M_{\rm{g}}>M_{\rm{g}}^{'}) \pm
                    \sqrt{N(M_{\rm{g}}>M_{\rm{g}}^{'})}]/N_{\rm{tot}},
  \label{eq:ppl2}
\end{equation}
where Poisson counting statistics were used to determine the uncertainty
in the number of disks larger than a given limit in the distribution
and $N_{\rm{tot}}=75$.
The probability determined from equation (\ref{eq:ppl2}) could be
assigned a corresponding metallicity, $Z^{'}$, from the relation
$M_{\rm{g}}^{'} = M_{\rm{g, crit}}$.
Equation (\ref{eq:mgcrit}) means that
\begin{equation}
  Z^{'} = -\log{0.01M_{\rm{g}}^{'}/M_{\rm{s, crit}}}.
  \label{eq:zdash}
\end{equation}
The value of $M_{\rm{s, crit}}$ was constrained to achieve
a mean planet probability for the metallicity range $Z=0.25-0.5$ dex
in agreement with that found by FV05, i.e., $P_{\rm{pl}} = 14.8 \pm 3.5$\%,
giving\footnote{The value derived in equation \ref{eq:mscrit} differs
slightly from the $0.24M_{\rm{J}}$ derived by Greaves et al. (2007) because
that paper included disks from AW05 of both class II and III in their
primordial gas mass distribution.}
\begin{equation}
  M_{\rm{s, crit}} = (0.5 \pm 0.1)M_{\rm{J}}.
  \label{eq:mscrit}
\end{equation}
The extrasolar planet-metallicity relation predicted by this model is
plotted in Fig.~\ref{fig:pplz2}, and shows good agreement with the
observed relation (equation \ref{eq:pplfv05}).

We have also inverted the problem by deducing the {\it required}
disk mass distribution that would lead to the solid line in Figure 
\ref{fig:pplz2} (i.e., $P_{\rm{pl}}(Z)$ parameterised according to
equation \ref{eq:pplfv05}).
In Figure \ref{fig:gasmass} we compare this required distribution
with the observed gas mass distribution.
Noting that this comparison can only be made over the upper quartile
of disk masses (since current planet detection statistics only extend to 
metallicities $<0.5$ and, in the model, it is only this range of disk
masses which can form planets in this metallicity regime), it is
evident that there is also good agreement between the model and observed 
distributions when plotted in this way.
To quantify this, we performed a one sided Kolmogorov-Smirnov test to compare
the distribution of gas masses inferred from AW05, when converted into
metallicity (equation \ref{eq:zdash}), with that inferred from equation
(\ref{eq:pplfv05}) for the range $Z^{'}=-0.5$ to 0.5 dex.
We found that discrepancies as large as or greater than those observed
occur in 69\% of samples of 75 members drawn
from a population with a cumulative distribution function in which
$P(Z<Z^{'}) = 0.03 \times 10^{2Z^{'}}$;
i.e., we conclude that the gas mass data are not unlikely to be drawn
from such a distribution, since at least 2 out of 3 times one would
expect data at least as discrepant as observed.

\section{Discussion}
\label{s:disc}
We have shown, under the assumption that a
critical solid mass in the protoplanetary disk is
required to form a planet that is detectable in radial
velocity surveys, that the observed frequency of planet
detections as a function of metallicity, $P_{\rm{pl}}(Z)$,
is compatible with the observed disk mass distribution
(as derived from sub-mm dust mass measurements of Classical
T Tauri stars in local star forming regions).
We now discuss the physical basis for this simple model and further
observational tests.

\subsection{Comparison with core accretion models}
\label{ss:ca}
\begin{figure}
  \begin{center}
      \vspace{-0.2in}
    \begin{tabular}{c}
      \hspace{-0.3in} \psfig{figure=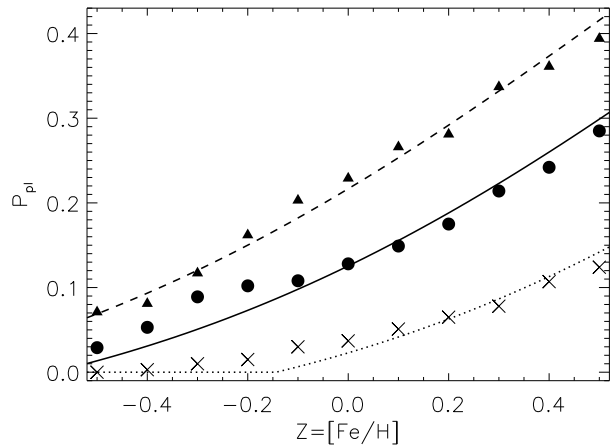,height=2.6in}
    \end{tabular}
    \caption{Prediction of the critical solid mass model
    for the probability that a star of given metallicity has an extrasolar
    planet that is detectable in the current radial velocity surveys assuming
    the disk mass distribution used in Ida \& Lin (2004b) with critical solid
    masses of 4 (dashed line), 8 (solid line) and 22 (dotted line) times the
    minimum mass solar nebula.
    The numerical results of the nominal model in IL04b are shown with filled circles,
    and the results of their models in which the core accretion rate is 3 times faster and
    slower than the nominal model shown with triangles and crosses (see their fig. 2b).}
    \label{fig:ppl4}
  \end{center}
\end{figure}

To consider the physical basis for the outcome of planet
formation being determined solely by dust mass,
we appeal to the core accretion models of Ida \& Lin (2004a, 2004b;
hereafter IL04).
The IL04 models are {\it local}, in the sense that planet
formation depends on local quantities such as gas and solid surface
density.
Therefore we expect any threshold effect to involve surface density
rather than mass.
We first assess whether the results of IL04 are compatible with the 
hypothesis that planet formation requires a critical {\it metallicity 
independent} solid surface density and return to a discussion of the 
relationship between solid surface density normalisation and dust mass in
\S \ref{ss:submm}.
We can assess this hypothesis in two ways.
Firstly, we can simply take the distribution of
disk surface densities assumed by IL04 (a log-normal distribution of
width 1.0 dex that is centred on the surface density of the minimum mass
solar nebula and truncated at $>1.48\sigma$), apply a
threshold solid surface density for planet formation
that is independent of metallicity and see whether we can
reproduce their numerical results.
Figure \ref{fig:ppl4} shows that this is indeed the case:
the nominal model from IL04b is well reproduced by assuming a critical
solid surface density of 8 times the minimum mass solar nebula,
whereas their variant models where the rate of core accretion is enhanced or
reduced by a factor of three are well reproduced by models in which the
critical solid surface density is respectively 4 and 22 times the minimum mass
solar nebula.
We stress that the IL04 models contain a large
number of ingredients and do not explicitly impose a threshold
criterion.
Nevertheless, we see that their results are empirically
equivalent to the imposition of a simple threshold.

In a second approach, we can now attempt to understand {\it why} the
IL04 models behave in this way.
Examination of these models shows that the formation of gas giant planets hinges on
rocky cores  being able to grow to a critical mass (a few $M_\oplus$) before the gas
disk is dispersed.
The requirement of sufficiently rapid core growth implies that they
have to form inside a critical radius, $a_{\rm{ig}}$, which depends on both
gas and solid surface densities.
On the other hand, inward of a second critical radius, $a_{\rm{tg}}$
(which depends on solid surface density), a critical core mass is not achievable 
because the required core mass exceeds the local isolation mass (at which point 
the core has consumed all the material in its local feeding zone).
Evidently, the formation of gas giant planets is possible only for the case
$a_{\rm{tg}} < a_{\rm{ig}}$ and we can derive a condition on the gas
and solid surface densities corresponding to the critical case where
$a_{\rm{tg}} = a_{\rm{ig}}$.
This translates into a condition on the minimum surface density of solids
as a function of metallicity.
We find that the critical surface density of solids scales as
$10^{-0.06 Z}$ (assuming, as in IL04, that a disk's surface density
scales $\Sigma \propto r^{-p}$ where $r$ is radius and $p=1.5$).
This very weak dependence on metallicity results from the
fact that the growth rate of solid cores is much more strongly dependent on
orbital radius than on the gas column density and hence $a_{\rm{ig}}$ is only
very weakly dependent on gas column density.
Therefore, the threshold criterion $a_{\rm{ig}} = a_{\rm{tg}}$ is nearly independent
of gas column density and thus the dependence of critical dust column on metallicity
is extremely weak.
It is this extremely weak dependence of the critical solid surface density
on metallicity which we believe to account for the excellent correspondence
between the numerical results of IL04 and the application of our
simple threshold hypothesis (see Fig.~\ref{fig:ppl4}).

A further test of this hypothesis would be to examine how $P_{\rm{pl}}(Z)$
predicted by the core accretion models depends on the assumed distribution of disk 
surface densities, since if the outcome is governed by a critical surface
density of solids for planet formation then using a narrower distribution of disk surface
densities as input would result in a steeper metallicity dependence (since in the
critical solid surface density model the metallicity dependence simply reflects the
disk surface density distribution used as input).
In contrast to IL04, Robinson et al. (2006) did vary this quantity and
indeed found that $P_{\rm{pl}}(Z)$ rose more gently when a
larger range of disk surface densities was employed.

\subsection{Why sub-mm dust mass determines outcome}
\label{ss:submm}
Regardless of the comparison with core accretion models, it is notable that
the critical solid mass model fits the planet-metallicity relation found in nature.
It is, however, surprising that sub-mm dust mass should be such a good
indicator of whether planets are going to form in a disk, since
sub-mm measurements probe the current mass in mm- to cm-sized dust
and so are not necessarily representative of the primordial inventory of
solid or gas mass.
Indeed, class II objects in Taurus-Auriga have a range of ages and
so we would expect the oldest stars to have already lost a significant quantity
of gas through accretion onto the star (Clarke et al. 2001).
We may also expect some loss of detectable dust mass with age through grain
growth and accretion onto the star with the gas.
However, there is no evidence that sub-mm dust mass changes with age
on the pre-main sequence (e.g., Wyatt et al. 2003) suggesting that the
mass in mm- to cm- sizes is constant.
This is to be expected, since the total dust mass $M_{\rm{dust}} \propto 
r_{\rm{out}}^{2-p}$, where $r_{\rm{out}}$ is the disk outer edge,
so that as long as $p < 2$ the sub-mm dust mass is concentrated in the outer
regions of the disk.
Since typically observed values for protoplanetary disks are $p \approx 0.85$
and $r_{\rm{out}} \approx 200$ AU (Andrews \& Williams 2007), the timescale for
grains containing most of the disk mass to grow to larger than 1 metre, and so
become invisible in the sub-mm, may be expected to be longer than the 10Myr
period over which planet formation (in the inner regions) must take place (e.g.,
Dullemond \& Dominik 2005).
Indeed some disks cannot harbour significant quantities of "unseen" dust mass
(i.e., with particle sizes either much larger or smaller than 1 mm),
since, even in the absence of such unseen contributions,
the gas mass inferred from mm dust measurements is in some cases already
$\sim 0.2$ times the central star's mass, and thus close to the limit
for gravitational instability.
Given the evidence that grain growth to mm and cm scales has occurred
in the outer regions of disks (Wilner et al. 2006), we are confident that this
grain size scale contains the majority of the disk solid mass at these radii,
and thus, by implication, the majority of the solid mass in the disk.
Thus, while the dust seen in the sub-mm is not
contributing to the planet formation process (because it is mainly at
radii where it has not had time to grow to large - greater than metre - 
size scales), we are suggesting that it is nevertheless a good measure
of the primordial inventory of solids in the disk.

The fact that the sub-mm dust mass distribution fits the observed planet-metallicity
relation so well is because there is an order of magnitude
difference between the highest and lowest masses
of the top $\sim 25$\% most massive gas disks
(e.g., Figs \ref{fig:gasmass} and \ref{fig:pplz2}).
This result is not specific to the Taurus-Auriga star forming region,
since class II disks in $\rho$ Oph also exhibit an order of magnitude
range for the most massive 25\% of those disks (see
Fig. 9 of Andr\'{e} \& Montmerle 1994).
If this distribution had been much narrower or broader
then we would have been able to rule out the critical solid mass
model.

One further requirement of nature for the critical solid mass model to work
is for a disk's outer radius to be less important than its solid mass
in setting the outcome of planet formation.
As noted in \S \ref{ss:ca}, models such as those in IL04 rely on a critical
surface density (rather than mass).
For the surface density profile assumed by IL04, the surface density
normalisation ($f_{\rm{d}}$, where $\Sigma \propto f_{\rm{d}}$), disk outer
radius ($r_{\rm{out}}$) and total solid mass ($M_{\rm{s}}$) are related via
$M_{\rm{s}} \propto f_{\rm{d}} r_{\rm{out}}^{0.5}$.
Thus the mapping between critical surface density and critical mass is
(weakly) dependent on $r_{\rm{out}}$.
While disk radii have been measured using sub-mm interferometry (Kitamura et al. 2002;
Andrews \& Williams 2007), these samples are biased toward the most
massive disks so that it is not clear how representative the observed distribution
is of the population as a whole.
However, there is no evidence that the distribution
of $r_{\rm{out}}$ is as broad as that of disk masses seen by AW05.
We therefore expect the surface density of solids in the planet formation region 
to be mainly controlled by $M_{\rm{s}}$ rather than $r_{\rm{out}}$, thus 
explaining the apparent success of sub-mm flux as a predictor of planet 
forming potential.

\subsection{Disks forming detectable planets}
\label{ss:detectable}
One implication of this study is that we can predict which
of the disks in the AW05 sample will go on to form
planets like those detected in the current radial velocity surveys.
The class IIs in their sample with more than $0.5M_{\rm{J}}$ of dust are
04113+2758, DL Tau, GG Tau and GO Tau.
However, we disqualify GG Tau as a planet-forming candidate, since its disk
is circumbinary (Guilloteau, Dutrey \& Simon 1999), and so its high sub-mm flux
does not equate with a high surface density of solids in the inner disk.
Massive circumbinary disks are rare (Jensen, Mathieu \& Fuller 1996),
so the majority of the more massive disks are not circumbinary disks
and so would not be unsuitable for forming planets.
Applying the same $0.5M_{\rm{J}}$ dust mass limit to the $\rho$ Oph study
of Andr\'{e} \& Montmerle (1994) indicates that of the class IIs in this
region, AS205, EL24, GSS39 and SR24S may go on to form detectable planets.

While we do not claim that we can unambiguously predict the outcome of planet
formation for any one of these systems, we do suggest that studying the
disks that are predicted to form planets, the characteristics of
which we can constrain at least statistically, may provide a valuable way of
probing the environments in which such planets form.
The fact that it is the most massive disks which go on to form detectable
planets means that these disks must be close to being gravitationally
unstable, since the ratio $M_{\rm{disk}}/M_{\star} > 0.05$ for $Z=0$ and 
$M_{\star}=1M_\odot$.
This suggests that instability could play a role in the formation process.
However, this cannot be the only determining factor, since the gravitational
instability process itself is not affected by metallicity (Cai et al. 2006),
and there would be no metallicity dependence if $M_{\rm{g, crit}}$ is a constant
and not dependent on metallicity.
Thus, this suggests that some degree of instability may help speed up
the core accretion process, e.g., through concentration of particles
in spiral structures (Rice et al. 2004) or instability in a
thin dust layer (Youdin \& Shu 2002).

\subsection{Observational tests}
\label{ss:ppl2}
Here we suggest three observational tests of the critical solid mass
model:

Firstly, if the model is correct, we would expect $P_{\rm{pl}}(Z)$ to
rise much less steeply with $Z$ at metallicities above $0.5$ dex
than implied by an extrapolation of equation (\ref{eq:pplfv05}),
since at higher metallicities the model predicts that planets
would be able to form in lower mass disks, and that $P_{\rm{pl}}(Z)$
in this regime would reflect the disk mass distribution of
intermediate mass disks.
A discrepancy between the observed disk mass distribution and that
resulting from an extrapolation of equation (\ref{eq:pplfv05})
to $Z>0.5$ dex is readily apparent by considering how the solid curve
on Figure \ref{fig:gasmass}, if extrapolated to lower disk masses,
would compare with the dashed line on that Figure.
Whether a suitable high metallicity sample can be found
to test this prediction remains to be seen (e.g., Laughlin 2000; Valenti \& Fischer 
2005; Taylor 2006).

Secondly, one of the key assumptions of the model was that the
distribution of protoplanetary disk masses is universal in that
it is independent of metallicity.
This can be tested by measuring the distribution of dust masses
in low (or high) metallicity star forming regions using sub-mm photometry,
since these masses should be correspondingly lower (or higher) than those of
nearby regions like Taurus-Auriga where $Z \approx 0$.
While ALMA can detect the brightest known class II disks out to
20 kpc, we are not aware of any young ($<10$ Myr) cluster within
the Milky Way which has a measured metallicity that is sufficiently
sub- or super-solar for the predicted difference in disk mass distribution
in comparison with Taurus-Auriga to be confidently detected, although 
star forming clusters such as those found by Santos et al. (2000) and
Yun et al. (2007) may be suitable candidates if their large
Galactocentric distances (15-16.5 kpc) are indicative of a low
metallicity as suggested by observations Cepheids which indicate a
metallicity gradient in the Milky Way of $-0.06$ dex/kpc (Luck et 
al. 2006).

\begin{figure}
  \begin{center}
      \vspace{-0.2in}
    \begin{tabular}{c}
      \hspace{-0.3in} \psfig{figure=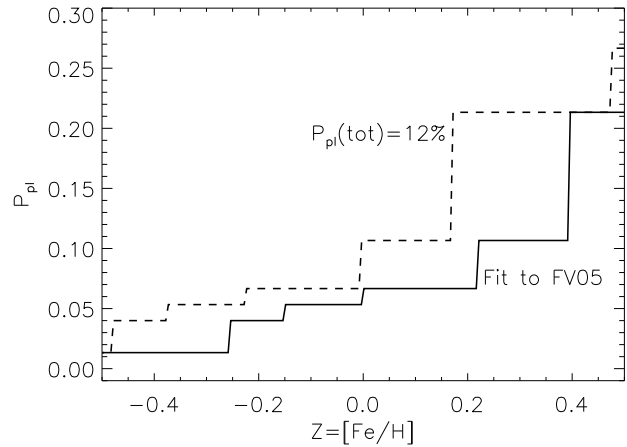,height=2.6in}
    \end{tabular}
    \caption{Prediction of the critical solid mass model
    for the probability that a star of given metallicity has an extrasolar
    planet.
    Assuming that surveys with different detection thresholds correspond
    to different critical solid masses, 
    the prediction for planets that are detectable in the current radial
    velocity surveys is shown with a solid line, and the prediction for
    a survey with an overall planet detection frequency of 12\% for
    the metallicity range -0.5 to 0.5 dex is shown with a dashed line.}
    \label{fig:pplz3}
  \end{center}
\end{figure}

Thirdly, we will be able to test in due course an adjunct hypothesis,
i.e., that the incidence of
planets of lower masses (and at greater orbital distances) is also
regulated by a (lower) critical solid mass threshold.
For example, extrapolation of the exoplanet
semi-major axis distribution to 20 AU suggests that surveys able to
detect planets to that distance would double the fraction of stars known
to have planets to 12\% (Marcy et al. 2005).
The simplest hypothesis we can apply to this population would simply be
that the progenitor disks corresponded to the top 12\%
of the disk mass distribution, implying a critical solid mass of
$\sim 0.3M_{\rm{J}}$.
We plot in Figure \ref{fig:pplz3} the predicted dependence of planet
frequency on metallicity in this case.
Although this "prediction" will eventually be compared with observational
data, we emphasise that it is not entirely clear how this adjunct
hypothesis (i.e., that the critical solid mass is lower for planets located
at larger distances) can be squared with the expectations of core accretion
models.

\section{Conclusions}
\label{s:conc}
We have presented a simple analytical model which can be used to predict the
outcome of planet formation, in which the formation of a planet that is detectable
in radial velocity studies depends only on the mass of solids in the
protoplanetary disk.
We showed that this model predicts that the observed planet-metallicity
relation is a reflection of the disk mass distribution.
We also argued that the sub-mm dust mass seen in the protoplanetary disk phase is a good tracer of 
the initial mass budget available close to the star for planet formation, and showed
that the observed planet-metallicity relation is consistent with the disk mass distribution
estimated from sub-mm observations of protoplanetary disks if the
critical solid mass required to form detectable planets is $0.5M_{\rm{J}}$.

We suggested that the detailed physics of the IL04 core accretion models
boil down to a critical solid mass required to form detectable planets, although
it needs to be confirmed that the good empirical agreement with the IL04 models
is more than a coincidence.
However, the value of this model is not just in its relevance to specific core
accretion models, but in its general applicability, since it shows how the
observed planet-metallicity relation would be reproduced by any
planet formation model which imposes a critical solid mass for the formation
of detectable planets.
Other reasons for imposing a threshold on a disk's solid mass before
detectable planets can form include the possibility that such
conditions are required for the formation of $>$ km-sized planetesimals
through gravitational instabilities (e.g., Johansen, Klahr \& Henning 2006).

\begin{figure}
  \begin{center}
      \vspace{-0.2in}
    \begin{tabular}{c}
      \hspace{-0.3in} \psfig{figure=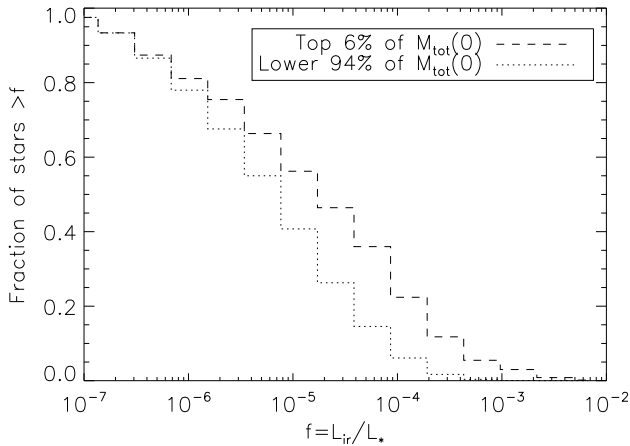,height=2.6in}
    \end{tabular}
    \caption{Distribution of infrared luminosities ($f=L_{\rm{ir}}/L_\star$)
    of the debris disks of A stars in the model of Wyatt et al. (2007).
    The distribution for the debris disks formed from the most massive 6\% of 
    protoplanetary disks (the \textit{planet bearers}) is compared with those 
    formed from the least massive 94\% of protoplanetary disks (the 
    \textit{non-planet bearers}).}
    \label{fig:astardisk}
  \end{center}
\end{figure}

The value of this model is also in its simplicity, since this means that it can be 
readily applied to predict other observable properties of stars with and without 
detectable planets, should those properties also depend on the solid mass of the 
protoplanetary disk.
For example, the statistics for the incidence of debris disks around A stars
as a function of stellar age (e.g., Rieke et al. 2005) can be explained by a
model in which all stars form planetesimal belts, the initial mass of which is
determined by the solid mass in the protoplanetary disk (a distribution which
is taken from AW05), and which are subsequently eroded by steady state collisional 
processing (Wyatt et al. 2007).
We ran the A star debris disk model in order to predict the distribution of
$f=L_{\rm{ir}}/L_\star$ for the ensemble, computing separate distributions
for the \textit{planet bearers} (corresponding to the top 6\% of the input
mass distribution) and the \textit{non-planet bearers} (corresponding to
the remaining 94\% of the population).
Fig.~\ref{fig:astardisk} shows how the debris disks of the \textit{planet bearers}
are, on average, more luminous than those of the \textit{non-planet bearers};
specifically, the mean luminosity of the most luminous 10\%
in both distributions (in $L_{\rm{ir}}/L_\star$) differ by a factor of
$\sim 6$.
While we do not know whether it is only the top 6\% of A star protoplanetary disks 
that form detectable planets, because the A star exoplanet population is poorly 
known at present, here we predict that if A star planets form in a similar manner to 
those of sun-like stars (i.e., with a threshold solid mass criterion), then this 
will be seen in the luminosity distributions of their debris disks 
(Fig.~\ref{fig:astardisk}).
Similarly, while we do not know whether the luminosity distributions of the
debris disks of sun-like stars behave as shown in Fig.~\ref{fig:astardisk},
because the model of Wyatt et al. (2007) has yet to be applied to that population, 
here we predict that if the luminosities of sun-like star debris disks are
governed by steady state processes, then the distributions of those luminosities
will exhibit a trend similar to that in Fig.~\ref{fig:astardisk}.
Indeed observations of the debris disks of sun-like stars both with and without 
planets do show a trend in their luminosities of comparable magnitude
to that suggested by Fig.~\ref{fig:astardisk} (Bryden et al. 2007).

Application of this model to known systems implies that the disks
of $04113+2758$, DL Tau, GO Tau, AS205, EL24, GSS39 and SR24S will form
(or have formed) gas giant planets.
While the outcome of planet formation in individual systems is uncertain,
we suggest that studying these disks may help constrain the physical
conditions of disks in which we know, at least statistically, what the outcome of
planet formation will be.
Observational tests of the model include a flattening of the metallicity relation
for $Z>0.5$ dex and also a flattening as planet searches continue.


\end{document}